\documentclass[preprint,showpacs,preprintnumbers,amsmath,amssymb, superscriptaddress,longbibliography,nofootinbib]{revtex4-1}
\usepackage{graphicx} 
\usepackage{xcolor}
\usepackage{amsmath}
 \usepackage{hyperref}
\usepackage[top = 2cm, bottom = 2cm, right = 2cm, left =2cm]{geometry}
\usepackage{orcidlink}
\begin{document}

\title{Non-Stationary Wormholes and Matter Source Reconstruction in General Relativity}

\author{G. Alencar \orcidlink{0000-0002-3020-4501}}
\email{geova@fisica.ufc.br}
\affiliation{Departamento de F\'isica, Universidade Federal do Cear\'a, Caixa Postal 6030, Campus do Pici, 60455-760 Fortaleza, Cear\'a, Brazil}

\author{T. M. Crispim \orcidlink{0009-0004-6604-4695}}
\email{tiago.crispim@fisica.ufc.br}
\affiliation{Departamento de F\'isica, Universidade Federal do Cear\'a, Caixa Postal 6030, Campus do Pici, 60455-760 Fortaleza, Cear\'a, Brazil}

\author{R. D\'{a}rlla \orcidlink{0009-0005-5203-8443}}
\email{robertadarlla2@gmail.com}
\affiliation{Departamento de F\'isica, Universidade Federal do Cear\'a, Caixa Postal 6030, Campus do Pici, 60455-760 Fortaleza, Cear\'a, Brazil}

\author{R. R. Landim \orcidlink{0000-0003-1387-5474}}
\email{renan@fisica.ufc.br}
\affiliation{Departamento de F\'isica, Universidade Federal do Cear\'a, Caixa Postal 6030, Campus do Pici, 60455-760 Fortaleza, Cear\'a, Brazil}

\date{\today}

\begin{abstract}
We investigate non-stationary wormhole geometries in General Relativity embedded in evolving cosmological backgrounds. We employ a Simpson-Visser-like regularization of the temporal coordinate of the $(n+1)$-dimensional FLRW metric, together with a homogeneous Cuscuton field, to construct cosmological backgrounds in which the scale factor remains finite and non-zero at the would-be initial singularity. We then consider two specific dynamical wormhole geometries: the four-dimensional defect wormhole and the topologically charged Perry-Mann type wormhole. We show that the former can be supported by a canonical scalar field with vanishing potential, while the latter requires a mimetic matter sector to support its spatial structure, with the Cuscuton governing the cosmological evolution in both cases. We subsequently generalize the analysis to a separable non-stationary spacetime and derive a scaling condition that strongly constrains the admissible matter sources. In particular, for minimally coupled scalar, this condition removes the functional freedom available in static reconstruction schemes. We show that, within this framework, regular, asymptotically flat wormhole geometries cannot be obtained while simultaneously satisfying the required smoothness and reflection symmetry conditions. Our results illustrate both the possibility of constructing specific non-stationary wormholes in regular cosmological backgrounds and the restrictions that arise when attempting to generalize such constructions to arbitrary spatial geometries.
\end{abstract}

\renewcommand{\thesection}      {\Roman{section}}
\maketitle
\newpage


\section{Introduction}

Over the past decades, General Relativity (GR) has proven to be a remarkably fertile framework for exploring non-trivial spacetime geometries. Among the most captivating theoretical possibilities are wormholes, hypothetical topological bridges connecting distant regions of the same universe or, potentially, distinct spacetimes through their throats. While initially conceptualized through the Einstein-Rosen bridge \cite{einstein1935particle}, traversable wormholes were brought into a more physically rigorous framework by the pioneering works of Ellis \cite{ellis1973ether}, Bronnikov \cite{bronnikov1973scalar}, and subsequently Morris and Thorne \cite{morris1988wormholes}. Traditionally, much of the literature has focused on static and spherically symmetric solutions \cite{Dai:2020rnc,Dai:2018vrw}. In these settings, sustaining a traversable throat typically requires "exotic matter", in the sense of matter that violates the usual energy conditions near the throat \cite{Odintsov2026}, most notably the null energy condition (NEC) \cite{visser1995lorentzian, hochberg1997geometric,Lobo:2020ffi}. Consequently, considerable effort has been devoted to identifying realistic physical mechanisms, alternative theories of gravity, or specific field couplings that can support such structures without relying on these violations.

Beyond their theoretical interest as exact solutions, wormholes have attracted renewed attention due to their potential observational signatures in the era of high-precision astrophysics \cite{Simonetti:2020ivl,Bambi:2021qfo}. Recent theoretical studies suggest that, depending on their geometric parameters and the nature of the matter supporting their throats, certain classes of wormholes can closely mimic the behavior of standard astrophysical black holes \cite{damour2007wormholes,Bambi:2013nla,Guerrero:2021ues,Azreg-Ainou:2014dwa,Ishkaeva:2023xny,alloqulov2024shadow,Nedkova:2013msa,Shaikh:2018kcg,Guerrero:2022qkh,Olmo:2021piq,Gyulchev:2018fmd,LimaJunior:2022zsp,Silva:2026koi,Crispim:2025cql,Abdikamalov:2019zr,He:2025acc,Nascimento:2025shadow}. For instance, wormhole spacetimes can exhibit gravitational lensing effects \cite{tsukamoto2012can,Nandi:2006ds,Tsukamoto:2020uay,Islam:2021ful,Pereira:2025fvg,Sharma:2022tiv,Sharma:2022dbx}, accretion disk emission profiles \cite{harko2008thin,Bambi:2013iron,Paul:2020acc,Crispim:2024dgd,Rueda:2023val,Combi:2024ehi}, and quasi-normal mode spectra (ringdown) \cite{cardoso2016gravitational, bueno2018echoes,Duran-Cabaces:2025sly,Churilova:2019qph,Volkel:2018hwb,Bronnikov:2019sbx,DuttaRoy:2019hij,Roy:2021jjg,kar1995resonances} that can closely resemble those of black holes, with deviations in some cases becoming significant only at late times. This phenomenological overlap motivates the study of regular, horizonless geometries as potential alternatives to black holes and as probes of strong-field gravity. Furthermore, their analysis can provide insight into the transition between regular black holes, black bounces, and traversable wormhole spacetimes \cite{simpson2019black,Alencar:2026ypy,Alencar:2025yyl,Silva:2025fqj,Pereira:2026ffn}.

On the other hand, cosmological solutions have also attracted considerable attention, particularly in the context of FLRW spacetimes. Embedding a wormhole into an expanding or contracting universe introduces a dynamical scale factor, allowing one to investigate how the throat geometry and its supporting matter are affected by cosmic evolution. This setting was explored in \cite{Alencar:2025nik}, where the study of dynamical wormholes in regular cosmological backgrounds raises the question of whether known wormhole solutions can be consistently extended to evolving spacetimes and, more importantly, what type of matter content is required to support them.

A particularly interesting example is the defect wormhole introduced by Klinkhamer, which describes a traversable wormhole without requiring exotic matter in the conventional sense. Instead, the geometry is associated with a spacetime defect characterized by a locally vanishing metric determinant \cite{Klinkhamer:2022rsj}. Another relevant example is the topologically charged Perry-Mann type wormhole, whose geometry incorporates non-trivial topological structure and has been studied in lower-dimensional gravitational settings \cite{Perryman}. These solutions provide useful test cases for investigating how specific wormhole geometries can be embedded into a cosmological background and whether their characteristic properties are preserved under such an extension.

A related possibility is provided by the Simpson-Visser regularization, which offers a prescription for constructing regular geometries from singular or non-trivial spacetime configurations \cite{Simpson:2018tsi}. Although originally introduced in the context of black holes and wormholes, this regularization can also be applied to the temporal coordinate of the FLRW metric, providing a simple way to construct nonsingular cosmological backgrounds. This approach was investigated in \cite{Alencar:2025nik}, and it provides the cosmological framework adopted in the present work.

In this work, we investigate the effects of applying the Simpson-Visser regularization to the temporal coordinate of the $(n+1)$-dimensional FLRW metric, with the aim of constructing a regular cosmological spacetime in which the scale factor does not vanish at the would-be initial singularity. We then construct cosmological extensions of the defect wormhole and the topologically charged Perry-Mann type wormhole. For each geometry, we analyze the corresponding Einstein field equations and characterize the matter content required to support the resulting configurations.

Having considered these two particular cases, we next ask to what extent this construction can be generalized to non-stationary spacetimes with more general spatial geometries. We therefore study a general separable non-stationary spacetime and derive the restrictions imposed by the Einstein equations on the matter sources capable of supporting it. This analysis allows us to distinguish the features that are generic to the cosmological extension from those that depend on the specific wormhole geometry and matter content.

This paper is organized as follows. In Sec.~\ref{sec:regular_cosmology}, we review the construction of regular FLRW cosmologies through a Simpson-Visser-like regularization of the temporal coordinate and the role of the Cuscuton field. In Sec.~\ref{sec:dynamic_wormholes}, we study the non-stationary geometries of the defect wormhole and the topologically charged Perry-Mann wormhole, together with their corresponding matter sources. In Sec.~\ref{sec:general_case}, we extend the analysis to a general separable non-stationary spacetime and derive the scaling conditions imposed on the matter sources by the Einstein equations. Finally, Sec.~\ref{sec:conclusions} summarizes our main results and discusses possible extensions.

\section{A Brief Review of Regular FLRW Cosmologies} \label{sec:regular_cosmology}

Before studying non-stationary wormhole geometries, it is important to first turn our attention to the Friedmann-Lemaître-Robertson-Walker (FLRW) model. In this framework, our Universe is described by a homogeneous and isotropic spacetime, whose metric is given by
\begin{equation}
    ds^2=-dt^2+a(t)^2\sum_{i,j=1}^n \delta_{ij}dx^idx^j,
\end{equation}
where $a(t)$ is the cosmological scale factor. The dynamics of this Universe is governed by the so-called Friedmann equations, obtained from Einstein's field equations in $(n+1)$ dimensions. For a spatially flat geometry $(k=0)$, they are given by
\begin{equation}
    \frac{n(n-1)}{2}H^2=\rho,
\end{equation}
and
\begin{equation}
    -(n-1)\dot{H} = \rho+p,
\end{equation}
where
\begin{equation}
    H\equiv \frac{\dot a}{a}
\end{equation}
is the Hubble parameter, $\rho$ represents the total energy density, and $p$ is the corresponding pressure. In this context, different matter sources may be considered, such as radiation, pressureless dust, dark energy, or scalar-field configurations, each leading to distinct cosmological evolutions.

Combining the Friedmann equations with the conservation equation $(\nabla_\mu T^{\mu\nu}=0)$
\begin{equation}
    \dot\rho+nH(\rho+p)=0,
\end{equation}
and assuming a barotropic equation of state
\begin{equation}
    p=w\rho,
\end{equation}
one finds that the scale factor evolves according to the power law
\begin{equation}
    a(t)\propto t^{\frac{2}{n(1+w)}}.
\end{equation}
For instance, in a radiation-dominated Universe, corresponding to
\begin{equation}
    w=\frac{1}{n},
\end{equation}
the scale factor behaves as
\begin{equation}
    a(t)\propto t^{\frac{2}{n+1}},
\end{equation}
while for pressureless dust $(w=0)$,
\begin{equation}
    a(t)\propto t^{\frac{2}{n}}.
\end{equation}

However, when the FLRW equations are extrapolated backwards in time, they generally lead to an initial spacelike singularity associated with the Big Bang, where physical quantities such as the energy density and curvature invariants diverge, and geodesic completeness breaks down \cite{Hawking:1970zqf, Penrose:1964wq}. In particular, the scale factor vanishes at $t=0$, implying that the physical volume of the Universe shrinks to zero and signaling the breakdown of the classical cosmological description in the high-energy regime.
In both cases, the scale factor vanishes at $t=0$, implying that the physical volume of the Universe shrinks to zero, whereas the energy density and spacetime curvature become singular. Such a pathological behavior indicates the breakdown of classical General Relativity in the high-energy regime and motivates the investigation of alternative cosmological scenarios capable of avoiding the primordial singularity. 

In order to investigate possible regular cosmological solutions, let us consider the cuscuton scalar field $\phi_1$ coupled to gravity in $(n+1)$ dimensions \cite{Afshordi:2006cg}. The action for the cuscuton sector is given by
\begin{equation}
    S_{\phi_1}=\int d^{n+1}x \sqrt{-g}\left[\mu\sqrt{2X}-U(\phi_1)\right],
\end{equation}
where $\mu$ is a constant parameter and
\begin{equation}
    X=-\frac12 g^{\mu\nu}\partial_\mu\phi_1\partial_\nu\phi_1 .
\end{equation}
Throughout this work, we assume a homogeneous scalar configuration,
\begin{equation}
    \phi_1=\phi_1(t),
\end{equation}
so that $X>0$ and the scalar gradient is time-like.

Varying the action with respect to the scalar field yields the cuscuton equation of motion,
\begin{equation}
    \nabla_\mu\left(
    \frac{\partial^\mu\phi_1}{\sqrt{2X}}
    \right)
    -\frac{1}{\mu}\frac{dU}{d\phi_1}=0 .
\end{equation}
On the other hand, variation with respect to the metric provides the corresponding energy-momentum tensor,
\begin{equation}
    T_{\mu\nu}
    =
    \mu\frac{\partial_\mu\phi_1\partial_\nu\phi_1}{\sqrt{2X}}
    +g_{\mu\nu}
    \left(
    \mu\sqrt{2X}-U(\phi_1)
    \right).
\end{equation}

For a homogeneous FLRW background, the nonvanishing mixed components of the energy-momentum tensor reduce to
\begin{equation}
    T^\mu{}_\nu=
    \mathrm{diag}
    \left(
    -U(\phi_1),
    \mu|\dot\phi_1|-U(\phi_1),
    \ldots,
    \mu|\dot\phi_1|-U(\phi_1)
    \right),
\end{equation}
where the spatial pressure term appears $n$ times. Therefore, the effective energy density and pressure associated with the cuscuton field are respectively given by
\begin{equation}
    \rho_{\phi_1}=U(\phi_1),
\end{equation}
and
\begin{equation}
    p_{\phi_1}=\mu|\dot\phi_1|-U(\phi_1).
\end{equation}
It is important to emphasize that the cuscuton field does not behave as a standard propagating scalar degree of freedom. Since the action is linear in the field velocity, the scalar equation of motion does not provide an independent dynamical evolution for $\phi_1$. Instead, the cuscuton sector acts as an effective geometrical component whose configuration is entirely determined by the cosmological background through the Friedmann equations. Consequently, the vacuum cuscuton system alone is generally insufficient to uniquely determine a regular cosmological evolution \cite{Quintin:2019orx, Bernardo:2022cck}.

To obtain regular configurations, the authors of Ref.~\cite{Alencar:2025nik} adopted a Simpson-Visser-like regularization procedure (See Ref.~\cite{Simpson:2018tsi}) in the temporal coordinate, namely
\begin{equation}
    t \to \sqrt{t^2 + t_0^2},
\end{equation}
such that the metric is now written as
\begin{equation}
    ds^2 = -dt^2 + a^2\left(\sqrt{t^2 + t_0^2}\right) \sum_{i,j=1}^n \delta_{ij} dx^i dx^j.
\end{equation}
In this framework, the scale factor $a$ no longer vanishes identically at $t=0$; instead, it reaches a non-zero minimum, $a(t=0) = a_0 \neq 0$, thereby eliminating the initial singularity. To dynamically realize this geometry, one can consider a Cuscuton field coupled to a fluid background, such that the modified Friedmann equations read
\begin{align}
    \frac{n(n-1)}{2} H^2 &= \rho + U(\phi_1), \\
    -(n-1)\dot{H} &= \rho + p + \mu|\dot{\phi}_1|.
\end{align}

By employing a barotropic equation of state $p = w\rho$ alongside the energy-momentum conservation equation for the fluid, we find that the energy density evolves as
\begin{equation}
    \rho(t) = \frac{\rho_0}{a^{n(1+w)}},
\end{equation}
which implies
\begin{align}
    U(\phi_1) &= \frac{n(n-1)H^2}{2} - \rho_0 a^{-n(1+w)}, \\
    \mu|\dot{\phi}_1| &= -(n-1)\dot{H} - \rho_0(1+w) a^{-n(1+w)}.
\end{align}
This system allows us to determine the field functions given a specific scale factor. As a concrete example, considering pressureless dust ($w=0$) and a bounce-like scale factor defined by
\begin{equation}
    a(t) = a_0 (t^2 + t_0^2)^{1/n},
\end{equation}
we obtain the following explicit analytical solutions:
\begin{align}
    \mu(\phi_1(t) - \phi_0) &= -\frac{2(n-1)}{n} \left[ \frac{t}{t^2 + t_0^2} + \frac{\rho_0 n}{2 t_0 a_0^n (n-1)} \arctan\left(\frac{t}{t_0}\right) \right], \\
    U(t) &= \frac{\frac{2(n-1)}{n} t^2 - \rho_0 a_0^{-n} (t^2 + t_0^2)}{(t^2 + t_0^2)^2}.
\end{align}
Therefore, within this framework, a regularized cosmological scenario can be constructed by allowing the Cuscuton field to coexist with matter. In what follows, we examine several examples of wormhole geometries embedded in a dynamic regular cosmological background \cite{Narahara:1994np, KordZangeneh:2025qel,Alencar:2025nik}.

\begin{figure*}[ht!]
\centering
{\includegraphics[scale=0.43]{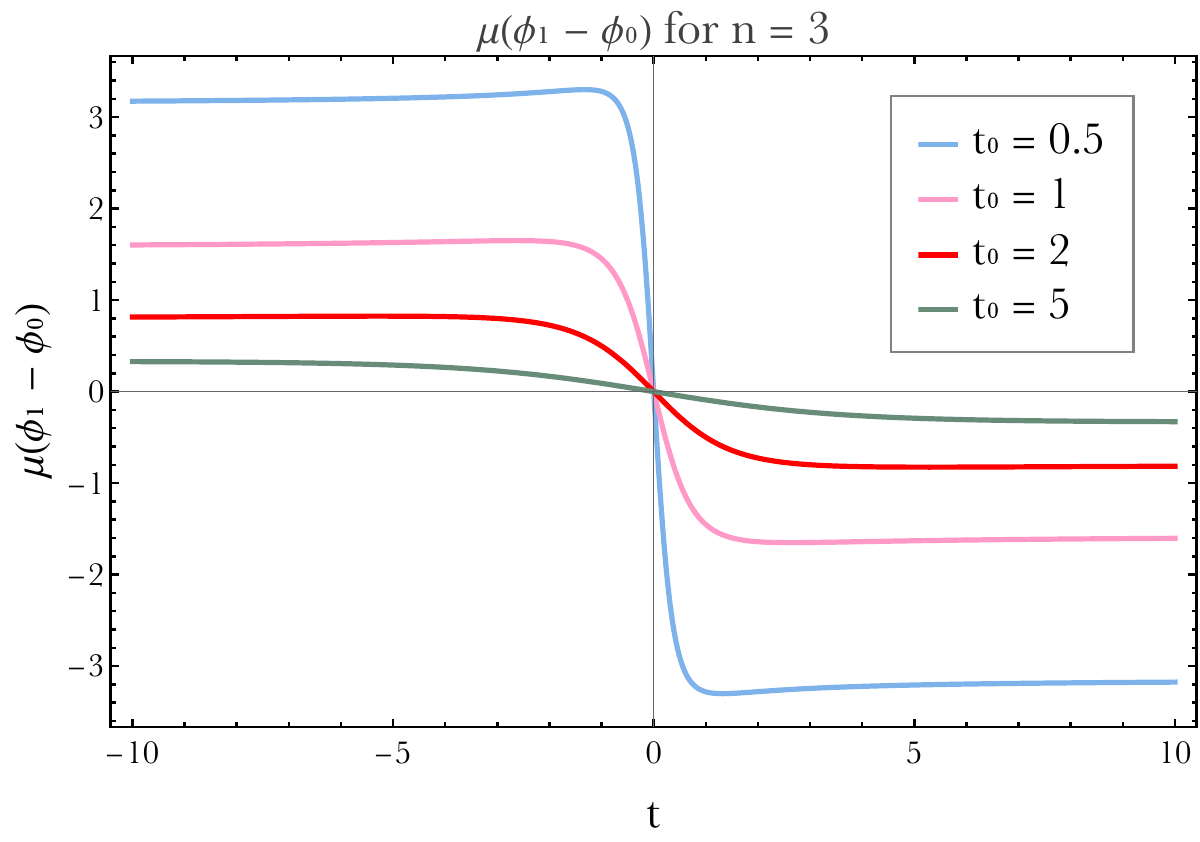}}
{\includegraphics[scale=0.43]{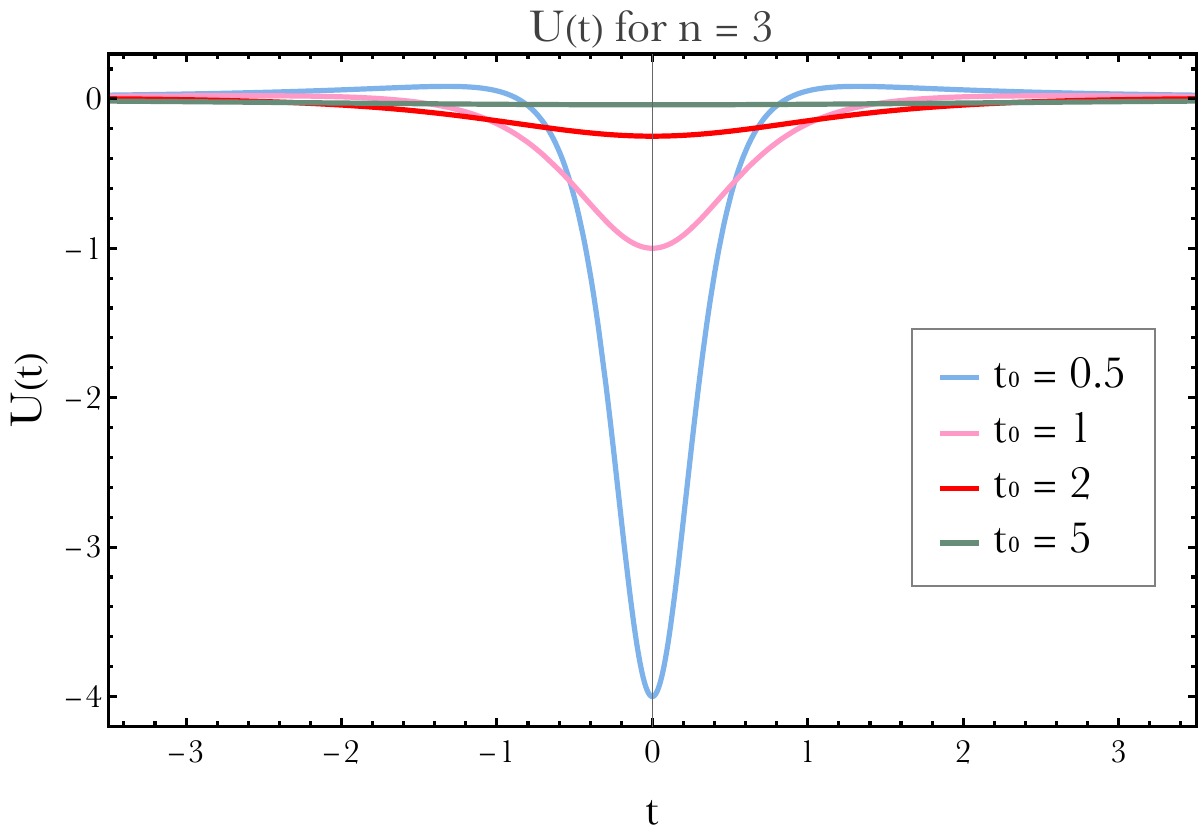}}
\centering \caption{Evolution of the cuscuton field $\phi_1$
and its potential $U(t)$ for different values of the
regularization parameter $t_0$. Here, $n=3$ and
$\phi_0=\rho_0=a_0=1$} \label{fig1}
{\includegraphics[scale=0.43]{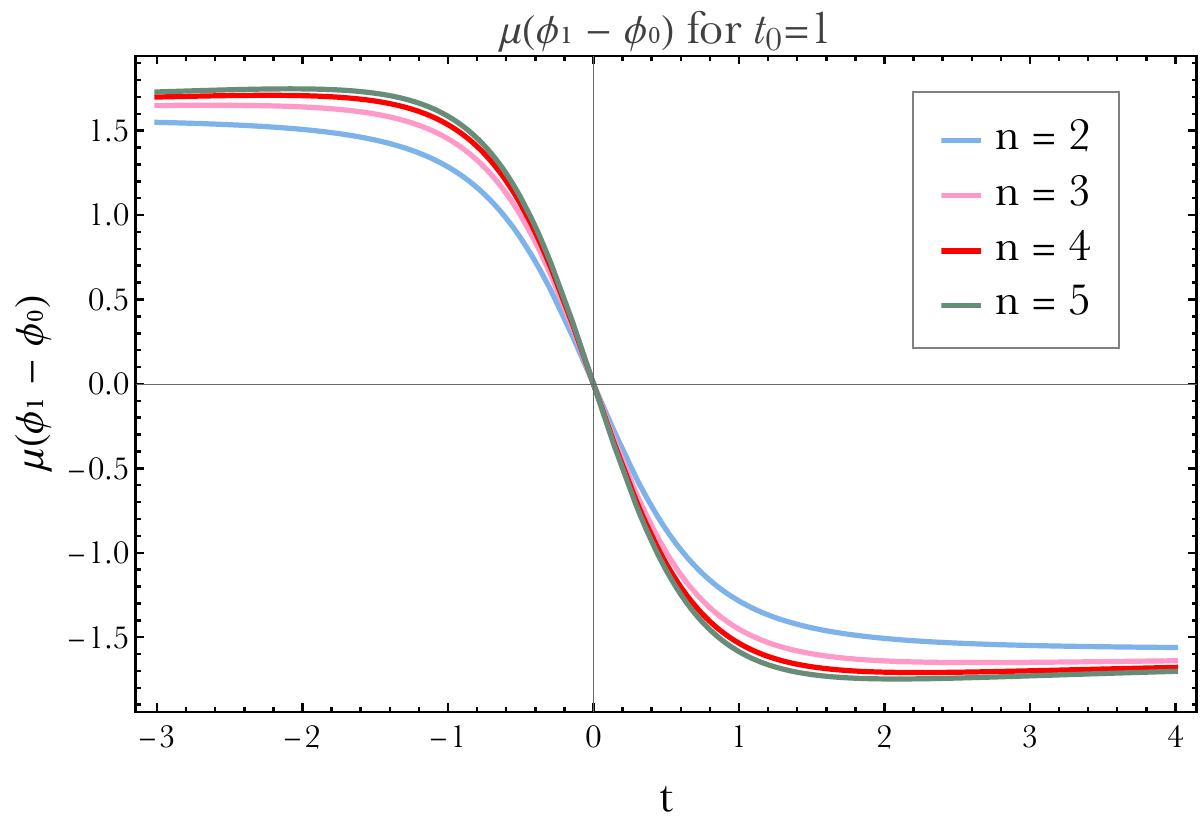}}
{\includegraphics[scale=0.43]{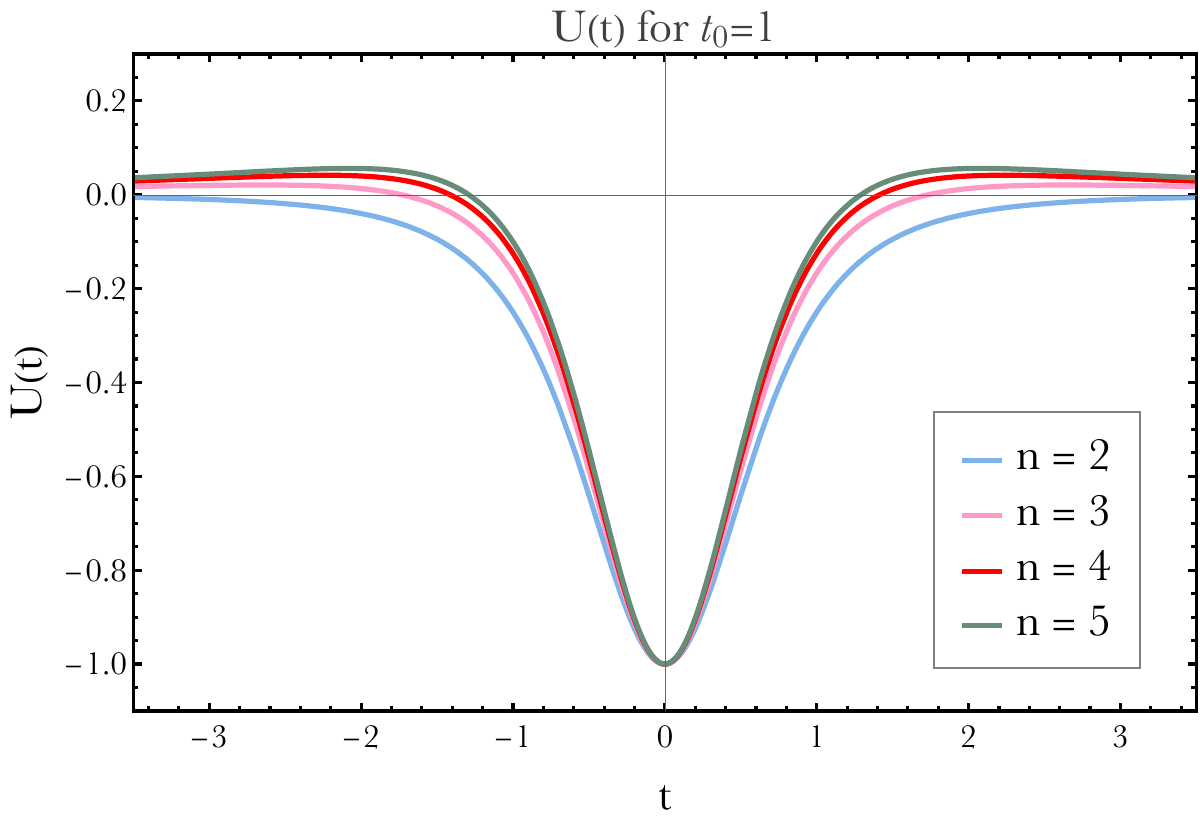}}
\centering \caption{Evolution of the cuscuton field
and its potential $U(t)$ for different values of the spatial
dimensionality parameter $n$. Here, $t_0=\phi_0=\rho_0=a_0=1$.} \label{fig2}
\end{figure*}

The behavior of the cuscuton field and its potential is illustrated in Fig. \ref{fig1} and Fig. \ref{fig2}. For fixed $n=3$, decreasing $t_0$ enhances and localizes the variation of the cuscuton field around $t=0$, while the magnitude of the potential at the bounce increases as $|U(0)|\propto t_0^{-2}$. Nevertheless, for any finite $t_0$, both $\mu(\phi_1(t)-\phi_0)$ and $U(t)$ remain finite at $t=0$. In the limit $|t|\rightarrow\infty$, the cuscuton field approaches $\mp\pi\rho_0/(2t_0)$, showing that the regularization parameter controls the asymptotic field amplitude as well as the temporal width of the transition. For fixed $t_0=1$, varying the number of spatial dimensions modifies the field profile and the behavior of the potential away from the bounce, while $U(0)=-1$ remains unchanged for the chosen parameters. In particular, $n=2$ is a special case for which the potential remains negative and approaches zero from below, whereas for $n>2$ the potential develops positive asymptotic tails and finite maxima away from the bounce. These results indicate that the regularization parameter primarily controls the strength and temporal scale of the regularized region, whereas the dimensionality affects the detailed structure of the cuscuton configuration away from the bounce.

\section{Examples of Dynamic Wormhole Geometries and Matter Sources In General Relativity} \label{sec:dynamic_wormholes}
As previously discussed, Ref. \cite{Alencar:2025nik} shows how to construct the field sources for the dynamic Ellis-Bronnikov wormhole, that is, one immersed in a regular cosmological background. To this end, the authors considered a free scalar field as the source supporting the throat, while the cosmological background is described by the framework outlined in the previous section. In this section, we will address specific examples of geometries for which we have successfully determined the dynamic field sources.

\subsection{Defect Wormholes: Traversability without Exotic Matter}

In the theoretical study of traversable wormholes, one of the most formidable physical obstacles is the standard requirement of exotic matter-sources that violate the Null Energy Condition (NEC) in order to keep the wormhole throat open. To circumvent this issue, the authors in Ref.~\cite{Klinkhamer:2022rsj} introduced the concept of a "defect wormhole". The line element for this wormhole can be written as:
\begin{equation}
    ds^2 = -dt^2 + \left(\frac{\xi^2}{\lambda^2+\xi^2} \right) d\xi^2 + (b_0^2+\xi^2)\,d\Omega_2^2,
\end{equation}
where $b_0$ and $\lambda$ are strictly positive constant parameters and $d\Omega_2^2=d\theta^2 + \sin^2\theta\, d\phi^2$. 
We can generalize their spatial geometry to our dynamic cosmological framework, we consider the following time-dependent ansatz:
\begin{equation}
    ds^2 = -dt^2 + a^2(t) \left(\frac{\xi^2}{\lambda^2+\xi^2} \right) d\xi^2 + a^2(t)(b_0^2+\xi^2)\,d\Omega_2^2.
\end{equation}
The most remarkable feature of this spacetime is its behavior at the throat, located at $\xi=0$. At this hypersurface, the spatial determinant of the metric vanishes, which introduces a \textbf{spacetime defect} rather than a standard curvature singularity. Because the defect is purely metric and topological in nature, all curvature invariants remain completely finite and smooth everywhere across the spacetime.

What makes this configuration particularly compelling from a physical standpoint is that the throat is sustained by this geometric defect instead of an unphysical fluid. As a result, it evades standard topological theorems that demand exotic matter. For the parameter range $\lambda^2 > b_0^2$, the model yields a strictly positive energy density and fully satisfies the NEC. Furthermore, in the critical case where $\lambda^2 = b_0^2$, the matter content vanishes identically, yielding an exact \textbf{vacuum wormhole} ($T_{\mu\nu}=0$). These well-behaved properties make the defect wormhole a highly attractive setup for exploring physically viable cosmological topologies.

The components of Einstein tensor for this geometry is
\begin{align}
G^{t}{}_{t} &= \frac{1}{a(t)^2}\frac{b_0^2 -\lambda^2}{(\xi^2 +b_0^2)^2}-\frac{3 \Dot{a}(t)^2}{a(t)^2},
\\[10pt]
G^{\xi}{}_{\xi} &=\frac{1}{a(t)^2}\frac{\lambda ^2-b_0^2}{ \left(\xi ^2+b_0^2\right)^2} -2\frac{\ddot{a}(t)}{a(t)}
-\left(\frac{\dot{a}(t)}{a(t)} \right)^{2} ,
\\[10pt]
G^{\theta}{}_{\theta} &= 
G^{\varphi}{}_{\varphi} =\frac{1}{a(t)^2}\frac{b_0^2-\lambda ^2}{ \left(\xi ^2+b_0^2\right)^2} -2\frac{\ddot{a}(t)}{a(t)}
-\left( \frac{\dot{a}(t)}{a(t)} \right)^{2} .
\end{align}

Another interesting aspect of this geometry is the fact that, like the Ellis-Bronnikov wormhole, it satisfies the symmetry typical of scalar fields, that is, $G^t_t = -G_\xi^\xi = G_\theta^\theta$. In order to reconstruct the corresponding matter Lagrangian that reproduces the dynamic defect wormhole solution, two scalar fields are considered, following the same procedure established for the Ellis-Bronnikov case. The action is given by
\begin{equation}
    S = \int d^4x \sqrt{-g} \left[ -\frac{\epsilon}{2} g^{\mu\nu}\partial_\mu\phi_2\partial_\nu\phi_2 - V(\phi_2) + \mu\sqrt{-g^{\mu\nu}\partial_\mu\phi_1\partial_\nu\phi_1} - U(\phi_1) \right].
\end{equation}
That is, the evolving defect wormhole will be described by the sum of these two scalar fields: the cuscuton field $\phi_1 = \phi_1(t)$ governing the cosmological background, and a standard scalar field $\phi_2 = \phi_2(\xi)$ supporting the wormhole throat. The parameter $\epsilon = \pm 1$ indicates whether the spatial field is canonical ($+1$) or a phantom ghost ($-1$).

For the cuscuton field $\phi_1 = \phi_1(t)$, the non-vanishing components of the energy-momentum tensor do not change from the standard cosmological case:
\begin{align}
    T^{(1)t}{}_{t} &= - U(\phi_1) , \\
    T^{(1)\xi}{}_{\xi} &= T^{(1)\theta}{}_{\theta} = T^{(1)\varphi}{}_{\varphi} = \mu\left|\dot{\phi}_1\right| - U(\phi_1).
\end{align}

Whereas the second scalar field $\phi_2 = \phi_2(\xi)$ is described by the energy-momentum tensor
\begin{align}
    T^{(2)t}{}_{t} &= -\frac{\epsilon}{2} \frac{\lambda^2+\xi^2}{a(t)^2\xi^2} (\phi_2')^2 - V(\phi_2) , \\
    T^{(2)\xi}{}_{\xi} &= \frac{\epsilon}{2} \frac{\lambda^2+\xi^2}{a(t)^2\xi^2} (\phi_2')^2 - V(\phi_2) , \\
    T^{(2)\theta}{}_{\theta} &= T^{(2)\varphi}{}_{\varphi} = -\frac{\epsilon}{2} \frac{\lambda^2+\xi^2}{a(t)^2\xi^2} (\phi_2')^2 - V(\phi_2) ,
\end{align}
where the prime denotes the derivative with respect to the spatial coordinate $\xi$.

By Einstein's field equations, 
\begin{equation}
    G^{\mu}{}_{\nu} = T^{(1)\mu}{}_{\nu} + T^{(2)\mu}{}_{\nu},
\end{equation} the general solution can be obtained by separating the radial and temporal parts. First, subtracting the angular equation from the radial equation ($G^{\xi}{}_{\xi} - G^{\theta}{}_{\theta} = T^{\xi}{}_{\xi} - T^{\theta}{}_{\theta}$), we obtain:
\begin{equation}
    \frac{2(\lambda^2-b_0^2)}{a(t)^2 (\xi^2+b_0^2)^2} = \epsilon \frac{\lambda^2+\xi^2}{a(t)^2\xi^2} (\phi_2')^2.
    \label{eq:spatial_separation}
\end{equation}
Notice that the scale factor $a(t)^2$ completely cancels out, allowing us to isolate the differential equation for $\phi_2$:
\begin{equation}
    \epsilon (\phi_2')^2 = \frac{2(\lambda^2-b_0^2)\xi^2}{(\lambda^2+\xi^2)(\xi^2+b_0^2)^2}.
\end{equation}

Here lies the most remarkable physical feature of the defect wormhole: since the geometry requires $\lambda^2 > b_0^2$ to configure a proper spacetime defect, the right-hand side of the equation is strictly positive. Consequently, we must fix $\mathbf{\epsilon = +1}$ for the scalar field to be real. In sharp contrast to the dynamic Ellis-Bronnikov wormhole, \textbf{this field is a canonical scalar field, not a physically inconsistent ghost}. 

To determine the scalar potential $V(\phi_2)$, we can evaluate the Klein-Gordon equation for the spatial field,
\begin{equation}
    \square\phi_2 - dV/d\phi_2 = 0.
\end{equation}
Using the Eq.~\eqref{eq:spatial_separation}, the d'Alembertian is given by
\begin{align}
    \square \phi_2 &= \frac{1}{\sqrt{-g}} \partial_\xi \left( \sqrt{-g} g^{\xi\xi} \phi_2' \right) \nonumber \\
    &\propto \partial_\xi \left[ \frac{\sqrt{\lambda^2+\xi^2}}{\xi} (\xi^2+b_0^2) \frac{\xi}{\sqrt{\lambda^2+\xi^2}(\xi^2+b_0^2)} \right] = \partial_\xi (1) = 0.
\end{align}
Since $\square \phi_2 = 0$ exactly, the Klein-Gordon equation strictly requires $dV/d\phi_2 = 0$. Setting the integration constant to zero, we naturally obtain a vanishing potential, $V(\phi_2) = 0$. 

With the potential uniquely determined, the differential equation for the canonical scalar field ($\epsilon=1$) can be analytically integrated, yielding a real logarithmic spatial profile:
\begin{equation}
    \phi_2(\xi) = \phi_0 + \frac{\sqrt{2}}{2} \ln \left| \frac{\sqrt{\lambda^2+\xi^2} - \sqrt{\lambda^2-b_0^2}}{\sqrt{\lambda^2+\xi^2} + \sqrt{\lambda^2-b_0^2}} \right|.
\end{equation}

\begin{figure*}[ht!]
\centering
{\includegraphics[scale=0.55]{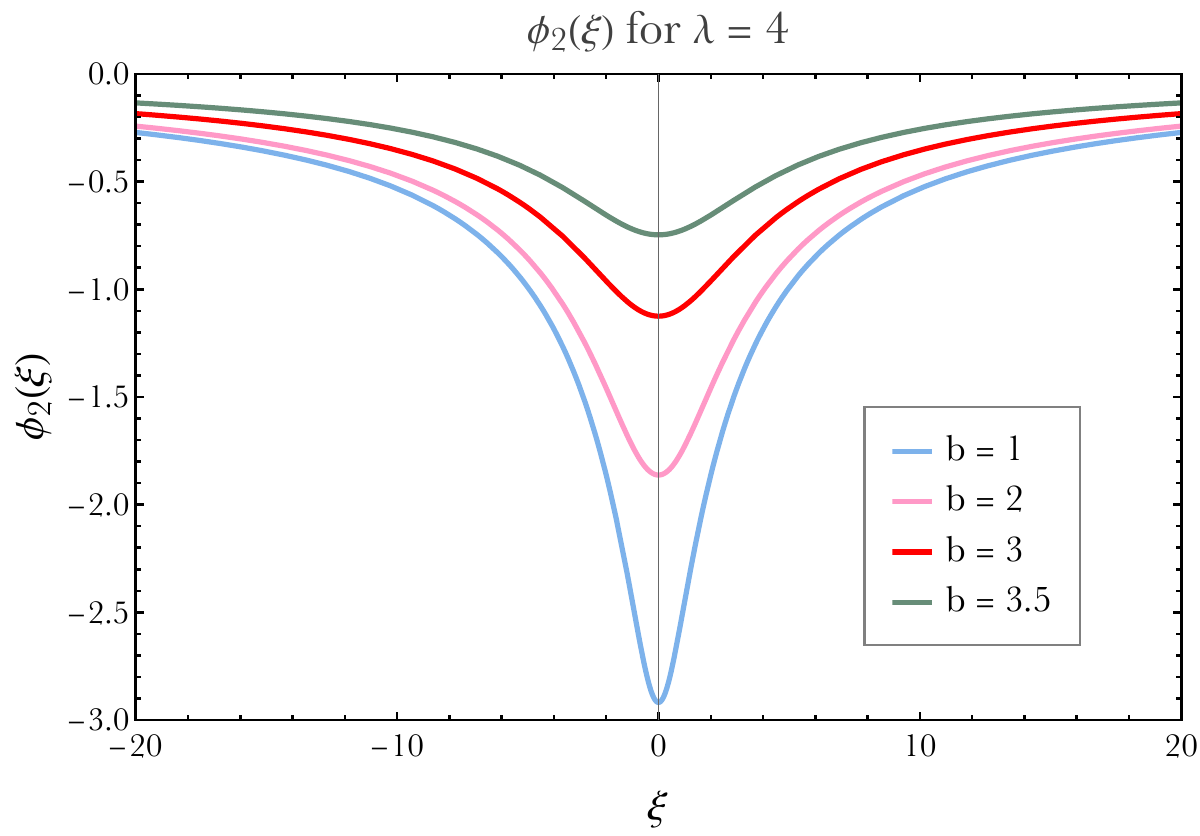}} \label{fig3}
\centering \caption{Scalar field $\phi_2(\xi)$ for different values of the wormhole parameter $b_0$. Here, $\lambda=4$ and $\lambda^2>b_0^2$.}
\end{figure*}
The behavior of the scalar field $\phi_2(\xi)$ is illustrated in Fig. \ref{fig3}. For fixed $\lambda=4$, varying the wormhole parameter $b_0$ modifies the depth and spatial extent of the scalar-field profile around the throat located at $\xi=0$. As $b_0$ increases, the magnitude of the field at the throat decreases, resulting in a progressively shallower profile. In particular, the case $b_0=1$ exhibits the largest deviation from the asymptotic value, whereas $b_0=3.5$ gives the least pronounced variation. For all considered values of $b_0$, the scalar field remains symmetric under $\xi\rightarrow-\xi$ and approaches $\phi_0$ as $|\xi|\rightarrow\infty$. This behavior follows from the dependence of the solution on $\sqrt{\lambda^2-b_0^2}$, with the condition $\lambda^2>b_0^2$ ensuring the reality of the scalar-field configuration. Thus, the parameter $b_0$ controls the amplitude of the scalar-field variation required around the wormhole throat, while $\lambda$ sets the characteristic scale of the configuration.

It is deeply insightful to analyze the limit $\lambda \to 0$ in this solution. Geometrically, taking $\lambda \to 0$ forces the radial metric component $g_{\xi\xi} = \xi^2/(\lambda^2+\xi^2)$ to smoothly approach unity ($g_{\xi\xi} \to 1$), which perfectly recovers the spatial structure of the Ellis-Bronnikov wormhole. However, dynamically, the quantity $\lambda^2 - b_0^2$ inside the square roots becomes negative, yielding an imaginary logarithmic argument. This mathematically reflects the physical necessity of a phantom ghost field ($\epsilon = -1$) to support the Ellis-Bronnikov geometry, which leads to an arctangent profile ($\arctan(\xi/b_0)$) instead. By keeping $\lambda > b_0$, the defect wormhole bypasses this constraint, guaranteeing a purely real solution supported exclusively by non-exotic matter ($\epsilon=1$).

Now, turning to the temporal side of the field equations, we substitute $\frac{1}{2}(\phi_2')^2$ and $V=0$ back into the $t-t$ component. The spatial geometric terms perfectly cancel each other out:
\begin{equation}
    G^{t}{}_{t} = \frac{1}{a(t)^2}\frac{b_0^2 -\lambda^2}{(\xi^2 +b_0^2)^2}-\frac{3 \Dot{a}(t)^2}{a(t)^2} = -U(\phi_1) + \frac{1}{a(t)^2}\frac{b_0^2-\lambda^2}{(\xi^2+b_0^2)^2}.
\end{equation}
This beautifully reduces to the standard Friedmann equation:
\begin{equation}
    3H^2 = U(\phi_1).
\end{equation}
Performing the equivalent substitution in the angular components yields the acceleration equation:
\begin{equation}
    -2\dot{H} - 3H^2 = \mu\left|\dot{\phi}_1\right| - U(\phi_1).
\end{equation}

As pointed out in previous sections, the temporal side implies that the cuscuton field equations reduce exactly to the standard cosmological framework. The only possible solution for an arbitrary potential is given by the roots $\phi_1 = \phi_0 = \text{constant}$ of the algebraic constraint equation:
\begin{equation}
    \left( \frac{dU}{d\phi_1} \right)^2 - 3\mu^2 U(\phi_1) = 0.
\end{equation}
For such a case, since $\dot{\phi}_1 = 0$, the Hubble parameter becomes a constant $H = \sqrt{U(\phi_0)/3}$. Consequently, the scale factor grows exponentially:
\begin{equation}
    a(t) \propto e^{\sqrt{\frac{U(\phi_0)}{3}} t}.
\end{equation}

This exponential expansion guarantees that the metric is entirely free of timelike singularities, describing a wormhole whose throat grows exponentially with time. Therefore, this framework successfully describes a regular, non-stationary defect wormhole embedded in a dynamic cosmology, fully supported by normal matter fields ($\epsilon=1$) and the geometric presence of the cuscuton, completely eliminating the necessity of exotic, ghost-like components.

\subsection{Topologically charged Perry-Mann type wormhole}

We consider the $(2+1)$-dimensional line element
\begin{equation}
    ds^{2} = -dt^{2} + a^{2}(t)\,dx^{2} + a^{2}(t)\,\beta^{2}\left(\alpha^{2}x^{2}+b^{2}\right)d\phi^{2},
\end{equation}
where $\alpha\leq 1$ controls the topological charge, $b$ is the throat parameter, and $\beta=1-4\lambda_d$ encodes the angular deficit associated with a spatial disclination. This geometry may be viewed as a cosmological generalization of the Perry-Mann wormhole, whose spatial profile is inherited from the function
\begin{equation}
    \Sigma(x)=\beta\sqrt{\alpha^{2}x^{2}+b^{2}}.
\end{equation}

For this spacetime, the non-vanishing mixed components of the Einstein tensor in $2+1$ dimensions are given by
\begin{align}
    G^{t}{}_{t} &= \frac{\alpha^{2}b^{2}}{a^{2}(t)\left(\alpha^{2}x^{2}+b^{2}\right)^{2}} - \frac{\dot{a}^{2}(t)}{a^{2}(t)}, \\[6pt]
    G^{x}{}_{x} &= G^{\phi}{}_{\phi} = -\frac{\ddot{a}(t)}{a(t)}.
\end{align}
Notice that the time-dependent cosmological evolution is entirely encoded in the scale factor $a(t)$ and its derivatives, whereas the topological wormhole profile appears exclusively in the purely spatial part of the $t-t$ component. 

To dynamically support this geometry, the field equations require an energy-momentum tensor with vanishing spatial pressures but a non-trivial, space-dependent energy density, effectively behaving as an inhomogeneous dust source. While one could phenomenologically introduce an \textit{ad hoc} pressureless fluid to provide this energy density, such an approach lacks a fundamental scalar Lagrangian description. To bypass this issue and formulate a rigorous action principle, we introduce a mimetic scalar field $\chi$ constrained by $g^{\mu\nu}\partial_\mu\chi\,\partial_\nu\chi=-1$. The associated Lagrange multiplier naturally supplies the required dust-like density \cite{Chamseddine2013,Myrzakulov2016,Sheykhi2021}. Furthermore, to drive the cosmological background, we couple it with the cuscuton field $\phi_1$. The total action is thus written as
\begin{equation}
    S = \int d^{3}x\,\sqrt{-g}\left[ R + \mu\sqrt{-g^{\mu\nu}\partial_\mu\phi_1\,\partial_\nu\phi_1} - U(\phi_1) + \lambda\left(g^{\mu\nu}\partial_\mu\chi\,\partial_\nu\chi+1\right) \right],
\end{equation}
where $\mu$ is a constant parameter, $U(\phi_1)$ is the cuscuton potential, and $\lambda$ is the mimetic multiplier.

Assuming a homogeneous cuscuton field $\phi_1=\phi_1(t)$, its energy-momentum tensor takes the form
\begin{equation}
    T^\mu{}_\nu(\phi_1) = \mathrm{diag}\left( -U(\phi_1),\; \mu|\dot{\phi}_1|-U(\phi_1),\; \mu|\dot{\phi}_1|-U(\phi_1) \right).
\end{equation}
On the other hand, choosing the mimetic field to be synchronized with cosmic time, $\chi=t$, the constraint is trivially satisfied, and its energy-momentum tensor becomes
\begin{equation}
    T_{\mu\nu}^{(\chi)} = 2\lambda\,\partial_\mu\chi\,\partial_\nu\chi \quad \implies \quad T^\mu{}_\nu(\chi) = \mathrm{diag}\left(-2\lambda,\,0,\,0\right).
\end{equation}
The conservation equation for the mimetic sector, $\nabla_\mu(\lambda\partial^\mu\chi)=0$, implies that the multiplier scales as
\begin{equation}
    \dot{\lambda} + 2H\lambda = 0 \quad \implies \quad \lambda(x,t) = \frac{\lambda_0(x)}{a^{2}(t)},
\end{equation}
where $H(t) \equiv \dot{a}/a$ is the Hubble parameter, and $\lambda_0(x)$ is an arbitrary spatial function to be fixed by the Einstein equations.

Equating 
\begin{equation}
    G^\mu{}_\nu = T^\mu{}_\nu(\phi_1) + T^\mu{}_\nu(\chi),
\end{equation}
 the field equations naturally split into temporal and spatial sectors. By matching the spatial components ($G^{x}{}_{x} = T^{x}{}_{x}$), we obtain:
\begin{equation}
    -\left(\dot{H} + H^2\right) = \mu|\dot{\phi}_1| - U(\phi_1).
    \label{eq:PM_spatial}
\end{equation}
Meanwhile, the $t-t$ component yields:
\begin{equation}
    -H^2 + \frac{\alpha^{2}b^{2}}{a^{2}(t)\left(\alpha^{2}x^{2}+b^{2}\right)^{2}} = -U(\phi_1) - 2\lambda(x,t).
\end{equation}
We can decouple the homogeneous background cosmology from the topological defect geometry by assigning the temporal terms to the cuscuton field and the spatial terms to the mimetic field. This identification provides the relations:
\begin{align}
    U(\phi_1) &= H^2, \label{eq:PM_U_H2} \\[6pt]
    -2\lambda(x,t) &= \frac{\alpha^{2}b^{2}}{a^{2}(t)\left(\alpha^{2}x^{2}+b^{2}\right)^{2}}.
\end{align}

Substituting Eq.~\eqref{eq:PM_U_H2} back into the spatial equation \eqref{eq:PM_spatial}, the terms proportional to $H^2$ cancel out, leading to a much simpler background differential equation:
\begin{equation}
    -\dot{H} = \mu|\dot{\phi}_1|.
\end{equation}
Just as in the $3+1$-dimensional cases analyzed in previous sections, the most direct way to obtain a regular cosmological evolution devoid of initial singularities is to assume a constant cuscuton configuration, $\phi_1 = \phi_0 = \text{constant}$. This immediately implies $\dot{\phi}_1 = 0$, which forces the Hubble parameter to be a constant:
\begin{equation}
    \dot{H} = 0 \quad \implies \quad H(t) = H_0 = \sqrt{U(\phi_0)}.
\end{equation}
Consequently, the scale factor undergoes an exponential growth,
\begin{equation}
    a(t) \propto e^{H_0 t},
\end{equation}
ensuring that the spatial volume never shrinks to zero. 

Finally, the spatial profile of the wormhole is completely reconstructed and sustained by the mimetic field density, requiring
\begin{equation}
    \lambda_0(x) = - \frac{\alpha^{2}b^{2}}{2\left(\alpha^{2}x^{2}+b^{2}\right)^{2}}.
\end{equation}
Therefore, this $2+1$-dimensional non-stationary geometry is beautifully supported by a twofold mechanism: the cuscuton field provides the vacuum energy that drives the regular exponential expansion of the spacetime, while the mimetic scalar field supplies the exact inhomogeneous structural support required to keep the topologically charged Perry-Mann throat open.

\begin{figure*}[ht!]
\centering
{\includegraphics[scale=0.55]{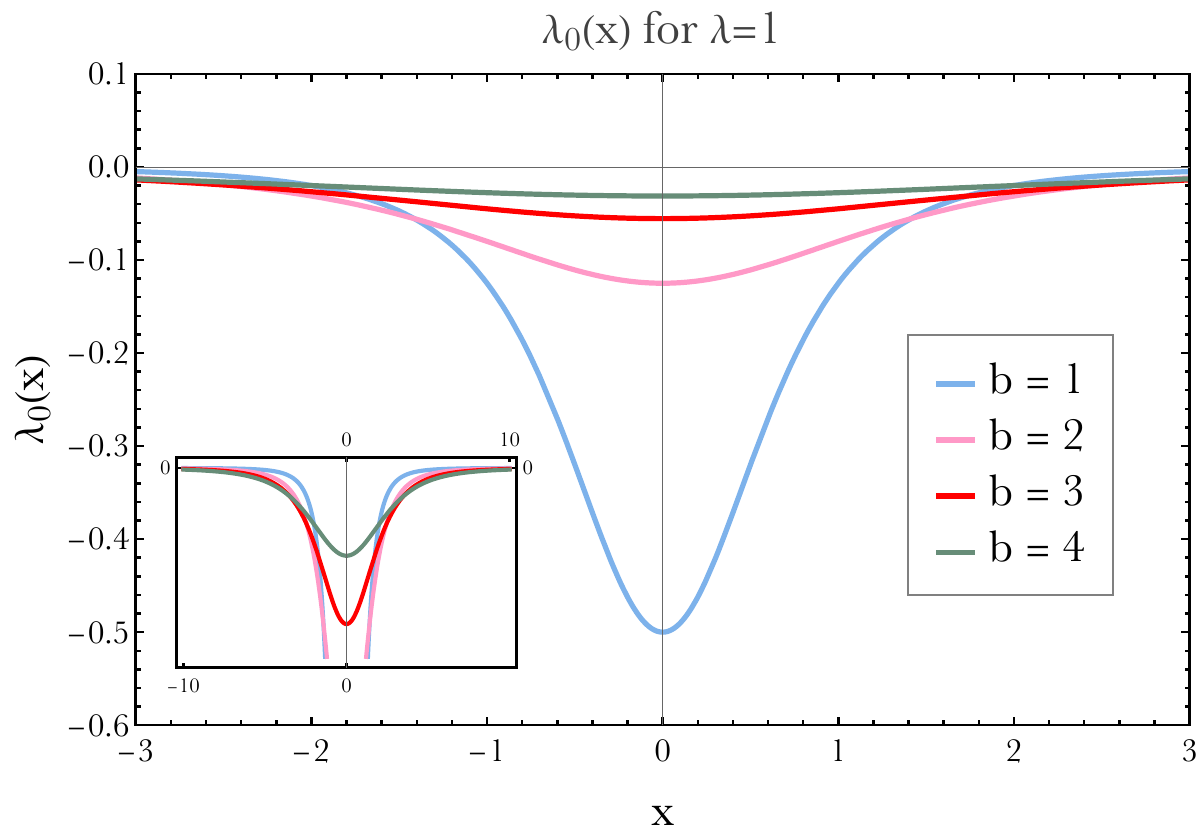}} \label{fig13}
\centering \caption{Mimetic field density $\lambda_0(x)$ for different values of the wormhole parameter $b$. Here, $\lambda=1$. The inset shows the same function over an extended spatial range.}
\end{figure*}

The behavior of the mimetic field density $\lambda_0(x)$ is illustrated in Fig. \ref{fig13} for different values of the wormhole parameter $b$, with $\lambda=1$. For fixed $\lambda$, increasing $b$ decreases the magnitude of the field density at the wormhole throat, $x=0$, while simultaneously broadening its spatial profile. In particular, the $b=1$ configuration exhibits the largest magnitude at the throat, whereas the variation becomes progressively weaker for larger values of $b$. For all considered cases, $\lambda_0(x)$ is symmetric under $x\rightarrow -x$ and approaches zero as $|x|\rightarrow\infty$, indicating that the mimetic contribution is localized around the wormhole throat. The inset displays the same function over an extended spatial range, emphasizing the strongly localized behavior of the field density near the origin and showing that all configurations become sharply peaked in the vicinity of $x=0$. These results show that the parameter $b$ controls both the amplitude and the spatial width of the mimetic field density, determining the distribution of the source required to sustain the wormhole geometry.

\section{The General Case} \label{sec:general_case}

We begin by considering a general non-stationary metric defined by the line element
\begin{equation}
    ds^2 = -dt^2 + a(t)^2 \bar{g}_{ij}(x) dx^i dx^j,
    \label{thegeneralcase}
\end{equation}
where $\bar{g}_{ij}$ (with $i,j=1,2,\dots,n$) represents an arbitrary time-independent spatial geometry. For this metric, the non-vanishing mixed components of the Einstein tensor $G^\mu{}_\nu$ naturally split into purely spatial terms scaled by the scale factor, and purely temporal terms:
\begin{align}
    G^0{}_0 &= \frac{\bar{G}^0{}_0}{a^2} - \frac{n(n-1)}{2} \frac{\dot{a}^2}{a^2}, \\[10pt]
    G^i{}_j &= \frac{\bar{G}^i{}_j}{a^2} - (n-1)\delta^i{}_j \left( \frac{\ddot{a}}{a} + \frac{(n-2)}{2} \frac{\dot{a}^2}{a^2} \right),
\end{align}
where $\bar{G}^\mu{}_\nu$ is the Einstein tensor associated exclusively with the static spatial metric $\bar{g}_{ij}$.

To support this evolving geometry, we assume a total matter action composed of a time-dependent cuscuton field $\phi_1(t)$-which will drive the background cosmological expansion-and a generic matter sector described by a Lagrangian $\mathcal{L}(\psi)$, where $\psi \equiv \psi(x^i)$ represents any set of spatial fields (scalar, vector, or tensor) responsible for sustaining the spatial geometry $\bar{g}_{ij}$. The general action reads
\begin{equation}
\label{acao_geral}
    S = \int d^{n+1}x \sqrt{-g} \left[ R + \mu \sqrt{-g^{\mu\nu} \partial_\mu \phi_1 \partial_\nu \phi_1} - U(\phi_1) + \mathcal{L}(\psi) \right],
\end{equation}
where $\mu$ is a constant parameter and $U(\phi_1)$ is the cuscuton potential.

For a homogeneous cuscuton field $\phi_1(t)$, its energy-momentum tensor is completely independent of the spatial coordinates, yielding
\begin{equation}
    T^0{}_0{}^{(\phi_1)} = -U(\phi_1), \qquad T^i{}_j{}^{(\phi_1)} = \delta^i{}_j \left( \mu |\dot{\phi}_1| - U(\phi_1) \right).
\end{equation}
By varying the action with respect to the metric, we obtain Einstein's equation.
\begin{equation}
  \label{einsteingeral}  G^\mu{}_\nu = T^\mu{}_\nu{}^{(\phi_1)} + T^\mu{}_\nu{}^{(\psi)}
\end{equation}
With this, we can establish a fundamental consistency condition for any matter field $\psi$ attempting to support this separable geometry. 

That is, if we have a non-stationary spacetime which described by the separable metric \eqref{thegeneralcase}, supported by a time-dependent cuscuton field $\phi_1(t)$ and a spatial matter sector $\psi(x^k)$, then the Einstein field equations are separable and consistent only if the energy-momentum tensor of the spatial generic field $\psi$ satisfies the dynamic scaling:
\begin{equation}
\label{consistency_theorem}
    T^\mu{}_\nu{}^{(\psi)} \propto \frac{1}{a(t)^2} \bar{T}^\mu{}_\nu{}^{(\psi)}(x^k).
\end{equation}
Consequently, any terms in $T^\mu{}_\nu{}^{(\psi)}$ that fail to scale identically with $a(t)^{-2}$, such as bare cosmological constants or unscaled interaction potentials, strictly break the separability of the field equations.

In fact, for the metric \eqref{thegeneralcase}, the non-vanishing mixed components of the Einstein tensor $G^\mu{}_\nu$ naturally decompose into spatial curvature terms $\bar{G}^\mu{}_\nu$, which scale exactly as $a(t)^{-2}$, and temporal terms depending solely on $a(t)$ and its derivatives. Because the components of $T^\mu{}_\nu{}^{(\phi_1)}$ depend exclusively on time, equating the field equations \eqref{einsteingeral} forces the spatial matter sector $T^\mu{}_\nu{}^{(\psi)}$ to absorb the $a(t)^{-2}$ scaling of $\bar{G}^\mu{}_\nu$. Any remaining unscaled terms in $T^\mu{}_\nu{}^{(\psi)}$ would introduce an anomalous time dependence into the spatial equations, rendering the system strictly inconsistent. 

As a concrete example, let us define $\psi$ as a spatial scalar field $\phi_2(x^i)$ with a self-interaction potential $V(\phi_2)$. The Lagrangian is given by
\begin{equation}
    \mathcal{L}(\phi_2) = \epsilon\, g^{\mu\nu} \partial_\mu \phi_2 \partial_\nu \phi_2 - V(\phi_2),
\end{equation}
where $\epsilon = \pm 1$ dictates whether the field is canonical ($+1$) or phantom ($-1$). Since $\phi_2$ depends only on the spatial coordinates, its energy-momentum components are evaluated as
\begin{align}
    T^0{}_0{}^{(\phi_2)} &= -\frac{\epsilon}{a(t)^2}\, \bar{g}^{ij}\partial_i\phi_2\partial_j\phi_2 + V(\phi_2), \\[6pt]
    T^i{}_j{}^{(\phi_2)} &= \frac{2\epsilon}{a(t)^2}\, \bar{g}^{ik}\partial_k\phi_2\partial_j\phi_2 - \delta^i{}_j \left[ \frac{\epsilon}{a(t)^2}\, \bar{g}^{kl}\partial_k\phi_2\partial_l\phi_2 - V(\phi_2) \right].
\end{align}
Applying our consistency condition \eqref{consistency_theorem}, it is immediately clear that the potential term $V(\phi_2)$ does not carry the necessary $a(t)^{-2}$ scaling. Therefore, to preserve the consistency of the field equations, the potential must strictly vanish, $V(\phi_2) = 0$. Consequently, the energy-momentum tensor drastically simplifies to:
\begin{align}
    T^0{}_0{}^{(\phi_2)} &= -\frac{\epsilon}{a(t)^2}\, \bar{g}^{ij}\partial_i\phi_2\partial_j\phi_2, \\[6pt]
    T^i{}_j{}^{(\phi_2)} &= \frac{\epsilon}{a(t)^2} \left( 2 \bar{g}^{ik}\partial_k\phi_2\partial_j\phi_2 - \delta^i{}_j \bar{g}^{kl}\partial_k\phi_2\partial_l\phi_2 \right).
\end{align}

This constraint is independently confirmed by the Klein-Gordon equation for the scalar field $\phi_2$. Varying the action yields
\begin{equation}
    \frac{1}{\sqrt{-g}} \partial_\mu \left( \sqrt{-g}\, 2\epsilon\, g^{\mu\nu} \partial_\nu \phi_2 \right) - \frac{dV}{d\phi_2} = 0.
\end{equation}
For the metric \eqref{thegeneralcase}, this becomes
\begin{equation}
    \frac{2\epsilon}{a(t)^2} \bar{\nabla}^2 \phi_2 - \frac{dV}{d\phi_2} = 0,
\end{equation}
where $\bar{\nabla}^2 = \frac{1}{\sqrt{\bar{g}}} \partial_i \left( \sqrt{\bar{g}}\, \bar{g}^{ij} \partial_j \right)$ is the Laplacian operator of the static spatial slice. Because $V$ is a function of $\phi_2(x^i)$ alone, the only way to satisfy this equation for all $t$ is, again, if $V = 0$ (modulo an irrelevant constant). Thus, the Klein-Gordon equation reduces simply to the Laplace equation:
\begin{equation}
    \bar{\nabla}^2 \phi_2 = 0.
\end{equation}

This result reveals a significant limitation. The scalar field $\phi_2$ is no longer a free dynamical degree of freedom; its configuration is entirely and rigidly determined by the spatial background geometry $\bar{g}_{ij}$ via the Laplace equation. If we assume $\bar{g}_{ij}$ describes a wormhole, the vanishing potential severely restricts the allowed geometries (e.g., essentially limiting the solutions to the Ellis-Bronnikov class, which inherently requires a phantom field, $\epsilon=-1$). 

On the other hand, by considering a special case, where the spatial metric represents a zero-tidal-force non-stationary geometry, the line element is given by
\begin{equation}
    \bar{g}_{ij}dx^idx^j = dr^2 + \Sigma^2(r)d\Omega_2^2,
\end{equation}
that its
\begin{equation}
    ds^2 = -dt^2 + a^2(t)\left[dr^2+\Sigma^2(r)d\Omega_2^2\right], 
\label{generalcase}
\end{equation}
where $\Sigma(r)$ is an arbitrary spatial function regulating the wormhole profile. For this ansatz, the non-zero components of the Einstein tensor are evaluated as
\begin{align}
    G^{t}{}_{t} &= \frac{1}{a^2(t)}\left(-\frac{1}{\Sigma^2} + \frac{\Sigma'^2}{\Sigma^2} + \frac{2\Sigma''}{\Sigma}\right) - 3\frac{\dot{a}^2(t)}{a^2(t)}, \\[6pt]
    G^{r}{}_{r} &= \frac{1}{a^2(t)}\left(-\frac{1}{\Sigma^2} + \frac{\Sigma'^2}{\Sigma^2}\right) - 2\frac{\ddot{a}(t)}{a(t)} - \frac{\dot{a}^2(t)}{a^2(t)}, \\[6pt]
    G^{\theta}{}_{\theta} &= G^{\varphi}{}_{\varphi} = \frac{\Sigma''}{a^2(t)\Sigma} - 2\frac{\ddot{a}(t)}{a(t)} - \frac{\dot{a}^2(t)}{a^2(t)}.
\end{align}

It is a well-established result in General Relativity that any static, spherically symmetric metric can be reconstructed as an exact solution to the Einstein equations by sourcing the geometry with a combination of a scalar field possessing a self-interaction potential and a magnetic field governed by a nonlinear electrodynamics (NED) Lagrangian $\mathcal{L}(\mathcal{F})$ \cite{Bronnikov:2022bud}. This powerful reconstruction method allows, for instance, the regularization of singular spacetimes into regular traversable wormholes. 

However, when we generalizing to non-stationary, separable geometries, this reconstruction method strictly breaks down. To establish this result, let us consider a separable, non-stationary, spherically symmetric spacetime be driven by a homogeneous cuscuton field $\phi_1(t)$ and spatially supported by a free scalar field $\phi_2(r)$ combined with a purely magnetic nonlinear electrodynamics (NED) source $\mathcal{L}(\mathcal{F})$. It is impossible to construct {\bf any general} smooth, reflection-symmetric, asymptotically flat wormholes within this framework.


The core mechanism underlying this reconstruction technique in static spacetimes relies on the presence of two arbitrary functional degrees of freedom in the matter sector: the self-interaction potential $V(\phi)$ and the nonlinear electrodynamics Lagrangian $\mathcal{L}(\mathcal{F})$. In a static metric, these two functions can always be engineered to match the spatial components of the Einstein tensor for any arbitrarily chosen profile $\Sigma(r)$.

In our non-stationary, separable framework, however, we define a rigid dynamical constraint: all spatial stress-energy components must scale strictly as $a(t)^{-2}$. This scaling requirement systematically eliminates both functional degrees of freedom:

\begin{enumerate}
    \item \textbf{Elimination of $V(\phi_2)$:} The potential $V(\phi_2)$ cannot absorb the required $a(t)^{-2}$ dynamic factor. Accordingly, this imposes a constraint $V(\phi_2) = 0$, stripping the scalar sector of its functional freedom and reducing $\phi_2$ to a free field governed by the spatial Laplace equation, $\bar{\nabla}^2\phi_2 = 0$. For the metric \eqref{generalcase}, this fixes $\phi_2'(r) = C_1/\Sigma(r)$.
    
    \item \textbf{Elimination of $\mathcal{L}(\mathcal{F})$:} For a static magnetic field configuration ($F_{\theta\varphi}=q_m \sin\theta$), the electromagnetic invariant scales as $\mathcal{F}(r,t) = \bar{\mathcal{F}}(r)/a^4(t)$. Requiring $T^\mu{}_\nu(\mathrm{NED}) \propto a(t)^{-2}$ forces the Lagrangian density to be uniquely proportional to the square root of the invariant:
    \begin{equation}
        \mathcal{L}(\mathcal{F}) = \beta\sqrt{\mathcal{F}},
    \end{equation}
    where $\beta$ is a constant. Consequently, $\mathcal{L}(\mathcal{F})$ ceases to be an adaptable functional degree of freedom and is pined down to a single mathematical form.
\end{enumerate}
With both $V(\phi_2)$ and $\mathcal{L}(\mathcal{F})$ completely fixed by dynamical consistency, the system loses the capacity to reconstruct an arbitrary geometry $\Sigma(r)$. 

Furthermore, for the square-root model $\mathcal{L}(\mathcal{F}) = \beta\sqrt{\mathcal{F}}$, the angular components of the NED energy-momentum tensor vanish identically ($T^\theta{}_\theta = T^\varphi{}_\varphi = 0$). Matching the angular component of the Einstein equations ($G^\theta{}_\theta = T^\theta{}_\theta$) then reduces the geometric freedom of the entire spacetime to a single, overconstrained ordinary differential equation:
\begin{equation}
\label{sigmanned1}
    \frac{\Sigma''}{\Sigma} = \frac{C_1^2}{2\Sigma^2} \quad \implies \quad \Sigma''\Sigma = \tilde{C},
\end{equation}
where $\tilde{C} \equiv C_1^2/2 \geq 0$ is a constant.

Integrating Eq.~\eqref{sigmanned1} with the throat boundary conditions $\Sigma(0)=\Sigma_0$ and $\Sigma'(0)=0$ yields:
\begin{equation}
    (\Sigma')^2 = 2\tilde{C} \ln\left(\frac{\Sigma}{\Sigma_0}\right).
\end{equation}
For any non-trivial scalar field ($\tilde{C} > 0$), as $\Sigma \to \infty$, the spatial derivative grows without bound as $\Sigma' \sim \sqrt{\ln\Sigma}$. Consequently, $\Sigma(r)$ grows strictly faster than linearly, violating the asymptotic flatness condition $\Sigma(r) \sim |r|$ as $|r| \to \infty$. 

If $\tilde{C} = 0$ (absence of the scalar field), Eq.~\eqref{sigmanned1} reduces to $\Sigma''=0$, whose reflection-symmetric solution is trivially constant ($\Sigma = \Sigma_0$), describing a cylindrical geometry rather than a wormhole. Therefore, the destruction of the functional degrees of freedom prevents Bronnikov-type metric reconstructions, prohibiting the existence of regular, asymptotically flat wormholes supported by this matter combination.

It is worth emphasizing that this no-go result that we established is strictly confined to the framework of standard General Relativity with minimally coupled matter fields. The breakdown of the geometric reconstruction method arises precisely because the minimal coupling isolates the spatial scaling constraint, rigidly fixing the available functional degrees of freedom ($V=0$ and $\mathcal{L} \propto \sqrt{\mathcal{F}}$).

However, if one considers non-minimal couplings between the scalar and electromagnetic sectors, for instance, via a generalized interacting Lagrangian of the form $\mathcal{L}(\phi_2, \mathcal{F})$, or extensions to modified gravity frameworks such as scalar-tensor theories, the theoretical landscape changes significantly. These extended scenarios naturally introduce additional, dynamically adaptable functional degrees of freedom that could potentially evade the restrictive scaling that we impose in the present work. Thereby reopening the theoretical possibility of reconstructing non-stationary, asymptotically flat wormholes.

%
%
%
%

\section{Final Remarks and conclusions} \label{sec:conclusions}

In this work, we have investigated non-stationary wormhole geometries embedded in regular cosmological backgrounds within General Relativity. Our analysis was motivated by the possibility of combining the regularization of the cosmological singularity with the construction of traversable wormhole geometries, while keeping the matter sources as physically consistent as possible. To this end, we employed a Simpson-Visser-like regularization of the temporal coordinate of the FLRW metric and considered a homogeneous Cuscuton field as the effective source responsible for the cosmological evolution.

We first reviewed the construction of regular FLRW cosmologies obtained through the replacement $t\rightarrow\sqrt{t^2+t_0^2}$. This prescription replaces the vanishing of the scale factor at the Big Bang by a non-zero minimum value and provides a nonsingular cosmological background. We showed that, when coupled to an additional matter sector, the Cuscuton field can consistently reproduce the required background dynamics. In particular, the explicit dust example illustrates how the Cuscuton potential and field configuration can be reconstructed for a regularized cosmological evolution. This provides the cosmological framework in which the dynamical wormhole geometries considered throughout the paper can be embedded.

We then investigated two specific examples of non-stationary wormholes. For the four-dimensional defect wormhole, we found that the spatial geometry can be supported by a real canonical scalar field with vanishing potential. In particular, the requirement $\lambda^2>b_0^2$ implies $\epsilon=+1$, showing that the dynamical defect wormhole does not require a phantom scalar to support its throat. The spatial scalar field is therefore qualitatively different from the phantom field required by the Ellis-Bronnikov geometry. The temporal sector, on the other hand, reduces to the Cuscuton cosmological equations, and a constant Cuscuton configuration leads to a constant Hubble parameter and consequently to an exponentially expanding scale factor. The resulting configuration provides a regular non-stationary defect wormhole supported by a canonical scalar field together with the cosmological Cuscuton sector.

A different mechanism emerges for the topologically charged Perry-Mann type wormhole in $(2+1)$ dimensions. In this case, the spatial dependence required by the wormhole geometry is naturally reproduced by a mimetic scalar field through its Lagrange multiplier, while the Cuscuton again governs the homogeneous cosmological sector. The Einstein equations separate into a temporal contribution associated with the cosmological background and a spatial contribution associated with the topological structure. For a constant Cuscuton field, the Hubble parameter is again constant and the scale factor evolves exponentially. Thus, although the two wormhole geometries require different matter sectors to support their spatial structure, they exhibit a common separation between the cosmological dynamics and the geometry of the wormhole throat.

Beyond these particular examples, we established a general consistency condition for non-stationary metrics of the form (\ref{thegeneralcase}). Since the spatial contribution to the Einstein tensor necessarily scales as $a^{-2}(t)$, any spatial matter sector responsible for supporting $\bar g_{ij}$ must possess the same temporal scaling. This requirement constitutes a strong restriction on the matter content compatible with separable non-stationary geometries. In particular, for a spatial scalar field, any self-interaction potential is incompatible with the required scaling and must therefore vanish, leaving a free scalar field satisfying the Laplace equation on the spatial slice.

We further applied this consistency condition to the reconstruction of general spherically symmetric wormhole geometries using a scalar field coupled to nonlinear electrodynamics. In static spacetimes, the scalar potential and the nonlinear electromagnetic Lagrangian provide functional degrees of freedom that can be adjusted to reconstruct a broad class of geometries. In the non-stationary separable setting considered here, however, the $a^{-2}$ scaling requirement removes this freedom: the scalar potential is forced to vanish, while the nonlinear electrodynamics Lagrangian is restricted to the square-root form $\mathcal{L}(\mathcal{F})\propto\sqrt{\mathcal{F}}$. The resulting angular Einstein equation then imposes an additional differential constraint on the wormhole profile. We showed that the corresponding solutions cannot simultaneously satisfy the requirements of smoothness, reflection symmetry, and asymptotic flatness for a non-trivial scalar field. In the absence of the scalar contribution, the remaining solution is cylindrical rather than wormhole-like. This establishes a no-go result for the construction of general regular, asymptotically flat non-stationary wormholes within this particular minimally coupled matter framework.

Taken together, these results reveal a fundamental distinction between specially constructed dynamic wormholes and arbitrary non-stationary wormhole geometries. Specific geometries, such as the defect and Perry-Mann type wormholes considered here, can be consistently embedded into an evolving cosmological background by appropriately identifying the matter sector responsible for the spatial structure. However, extending the reconstruction procedure to an arbitrary spatial geometry is strongly constrained by the temporal scaling imposed by the separable metric ansatz. Thus, the existence of regular dynamical wormholes is not determined solely by the regularity of the cosmological background; it also depends crucially on whether the matter sector retains sufficient functional freedom to reproduce the desired spatial geometry.

The no-go result obtained here is specific to General Relativity with minimally coupled matter fields and to the separable non-stationary framework considered in this work. It therefore does not exclude dynamical wormholes in more general gravitational or matter models. In particular, non-minimal couplings between scalar and electromagnetic fields, more general interacting matter sectors, or modified theories of gravity may provide additional functional degrees of freedom capable of circumventing the scaling restriction derived here. Investigating whether such extensions allow the reconstruction of regular, asymptotically flat non-stationary wormholes represents a natural direction for future work.

\section{ACKNOWLEDGMENTS}

We acknowledge the financial support provided by
the Conselho Nacional de Desenvolvimento Científico e
Tecnológico (CNPq), Fundação Cearense de Apoio ao
Desenvolvimento Científico e Tecnológico (FUNCAP).
This study was financed in part by the Coordenação de
Aperfeiçoamento de Pessoal de Nível Superior- Brasil
(CAPES)- Finance Code 001.

\bibliography{ref}

\end{document}